\documentclass[twocolumn,showpacs,preprintnumbers,amsmath,amssymb]{revtex4}

\usepackage[dvipdfm]{graphicx}
\usepackage{dcolumn}
\usepackage{bm}

\begin{document}

\preprint{APS/123-QED}

\title{Static Friction as a Function of Waiting Time\\ Probed by Dynamics of Driven Vortices in La$_{2-x}$Sr$_x$CuO$_4$ Thin Films}

\author{D. Nakamura}
\author{T. Kubo}%
\author{S. Kitamura}%
\author{A. Maeda}%
\affiliation{%
Department of Basic Science, the University of Tokyo, 3-8-1, Komaba, Meguro-ku, Tokyo, 153-8902, Japan}%


\date{\today}

\begin{abstract}
We investigated the dynamics of driven vortices in high-$T_c$ superconductor as an ideal model system to study the physics of friction.
The waiting-time dependence of the maximum static friction force, $F_s(t_w)$, was measured in La$_{2-x}$Sr$_x$CuO$_4$ thin films with different structures, sample sizes and pinning force.
We found various kinds of $F_s(t_w)$ in the $B$-$T$ phase diagram and in different types of samples.
The results suggest that the relaxation by thermal fluctuation is strongly affected by the pinning strength, the vortex bundle size and the system size.
Based on these results, we found crucial conditions to determine the validity of the Amontons-Coulomb's law, and proposed a criterion.
\end{abstract}

\pacs{Valid PACS appear here}
\maketitle

The physics of friction has attracted both scientists and engineers more than 500 years\cite{persson, bhushan}.
The empirical law of friction is known as the Amontons-Coulomb's (AC) law, describing (1) the kinetic friction force, $F_k$, is independent of the sliding velocity, $v$, (2) the maximum static friction force, $F_s$, is constant.
However, in reality, $F_k$ depends on the velocity, and $F_s$ depends on the waiting time, $t_w$, which is the intermission time of a repetitively applied driving force.
Thus, the AC law represents only very limited situations in the friction phenomena.
The deformation and the destruction of asperity of an interface\cite{bowden} and the motion of debris\cite{muser} are considered to be the candidates for the origin of such deviations from the AC law.
The criteria for the validity of the AC law is not simply determined by the surface roughness alone.
In fact, $F_s(t_w)$ was found not only in the friction at the macroscopic system such as the solid-solid interface of a massive block\cite{scholz} and the soft materials\cite{nitta}, but also in the friction at the lubricated microscopic interface, which is atomically smooth\cite{yoshizawa}.
Scientific challenges to elucidate the physics of friction have recently made progresses in some aspects\cite{yoshizawa,robbins}.
However, the comprehensive and microscopic understanding of the friction phenomena in any spatial scales and the hierarchy in the friction phenomena are still completely open issues.

We have utilized the dynamics of driven vortices in high-$T_c$ superconductors\cite{blatter} to investigate these unsettled subjects in the physics of friction\cite{maedainoue, maedanakamura}, since it has many common aspects to the physics of friction.
In the physics of friction, the object starts to move above $F_s$ and its dynamics has various characteristics, such as the stationary motion, the stick-slip motion, and the intermediate motion between them.
Even if $F<F_s$, thermal fluctuation causes microscopic relaxation of the interface because of the randomness and the multi-internal-degrees of freedom.
As a result, various kinds of aging effects and memory effects are often observed\cite{persson, bhushan}.
On the other hand, in the case of the dynamics of driven vortices, vortices can escape from the pinning potential in a sample by applying sufficient external driving current density above the critical value, $j_c$.
In addition, thermal fluctuation causes vortices to move with a finite net velocity even below $j_c$ and vortices relax into the more stable position.
As a consequence, the aging effects and the memory effects are commonly observed also in vortices in superconductors\cite{Du, Xiao, Henderson}.
Similar effects have been also observed in other quantum condensate, such as charge density wave\cite{ogawa} and domain walls in ferromagnet\cite{weger}.

A merit of our approach is that it does not contain either deteriorations or damages in a sample during experiments, so we can repeat experiments under the same external environment.
This is the significant advantage, because the measurement at the real microscopic interface is often influenced by the existence of wear, the junction growth by adhesion, and contamination materials, $etc$.
By measuring the  $I$-$V$ characteristics of driven vortices, explicit correspondence to the physical quantities shown up in the friction at the solid-solid interface has been obtained as follows\cite{matsukawa, maedainoue}.
The maximum static friction force, $F_s$, is equal to the Lorentz force at $j_c$; $F_s = j_c \Phi_0$, where $\Phi_0 = hc / 2e$ ($h$ is Planck constant, $c$ is the speed of light, and $e$ is electron charge) is the flux quantum.
On the other hand, we can obtain the kinetic friction force, $F_k$, by extracting the pinning force, which is the driving force minus the viscous drag in the steady state; $F_k = j \Phi_0 - \eta v$, where $\eta$ is the viscous drag coefficient.
We previously observed the strongly velocity-dependent $F_k$ for wide velocity range in high-$T_c$ superconductors La$_{2-x}$Sr$_x$CuO$_4$ and Bi$_2$Sr$_2$CaCu$_2$O$_y$, and we could scan various regions in the dynamic phase diagram of driven vortices from AC type to non-AC type by tuning the magnetic field, $B$, temperature, $T$, and also the pinning strength\cite{maedainoue, maedanakamura}.
This result suggests that the drastic change of the memory effect should be also observed in the same systems.
Furthermore, we expect that the criteria for the validity of the AC law will be obtained by comparing the results of $F_s(t_w)$ with $F_k(v)$.
Therefore, in this paper, another key phenomenon for the friction at the solid-solid interface, $F_s(t_w)$, was investigated in La$_{2-x}$Sr$_x$CuO$_4$ thin films.
By changing the strength of the pinning force by irradiating the columnar defects and changing the sample size by fabricating the bridge-type structure with different sizes, we aim to clarify the criteria which discriminates the AC type friction from the various different types of friction in real systems.


\begin{figure}[t]
\begin{center}
\hspace{0.5cm}
\includegraphics[width=0.9\linewidth]{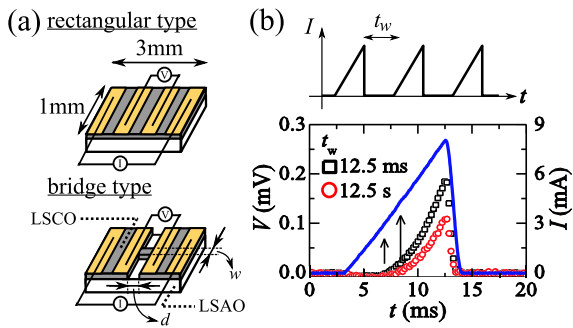}\\
\vspace{-0.2cm}
\caption{\label{label}(a) Typical shapes of thin films with the rectangular-type structure and with the bridge-type structure. (b) (Top panel) Schematic figure of the repetitively applied driving current as a function of time.
(Bottom panel) Raw data of the transient response of vortices.
Solid line is an applied driving current, and open symbols are the induced voltages in a sample with different waiting times, $t_w$'s. Arrows represent the appearance of finite voltages.} 
\end{center}
\end{figure}

\begin{center}
\begin{table}[t]
\centering
\caption{Sample profiles. The nominal carrier concentration, $x$, the width, $w$, and the distance, $d$, between voltage electrodes and the superconducting transition temperature, $T_c$, are shown.
R, B, and C represent the rectangular-type structure, the bridge-type structure, and the bridge-type structure with columnar defects, respectively.}
\label{units}
\begin{tabular}{c|c|c|c|c}
 & $x$ & $w$ ($\mu$m) & $d$ ($\mu$m) & $T_c$ (K) \\
\hline
\#R1 & 0.12 & 1470 & 660 & 35.20 \\
\#R2 & 0.15 & 1180 & 210 & 35.06 \\

\#B1 & 0.15 & 11.1 & 51.6 & 35.89 \\
\#B2 & 0.15 & 43.8 & 87.5 & 38.91 \\
\#B3 & 0.15 & 95.1 & 110 & 28.76 \\

\#C1& 0.15 & 40.0 & 90.0 & 35.11 \\
\end{tabular}
\end{table}
\end{center}

\vspace{-1cm}
Films of the underdoped ($x$=0.12) and the optimally doped ($x$=0.15) La$_{2-x}$Sr$_{x}$CuO$_4$ with 3000 \AA  -thickness on LaSrAlO$_4$ substrate were prepared by the pulsed laser deposition technique\cite{tsukada}.
Three types of thin films were prepared (Fig. 1(a)).
The 1st type is the thin films with the rectangular-type structure (\#R1, \#R2).
The 2nd type is the thin films with the bridge-type structure (\#B1$\sim$\#B3), fabricated by the photolithography and the chemical etching technique.
The 3rd type is the bridge-type thin film with the columnar defects (\#C1), introduced by the irradiation of 5.8 GeV Pb ions parallel to the $c$-axis using the Grand Acc\'el\'erateur National d'Ions Lourds (GANIL) in Caen, France.
Electrodes were made by painting Au paste.
The nominal carrier concentration, $x$, the width, $w$, and the distance, $d$, between the voltage electrodes, and the superconducting transition temperature, $T_c$, are shown in Table I.

Magnetic fields were applied along the $c$-axis with field-cooled conditions to avoid possible non-uniformity of vortex density.
Also, we waited about one hour before each experiment to achieve homogeneous distribution of vortices.
A sawtooth-like pulsed electrical current with $t_w$ was applied to obtain the transient response of vortices, as is shown in the top panel of Fig. 1(b).
Bottom panel of Fig. 1(b) shows the raw data of the experiments.
Solid line is driving current, and open symbols are waveforms of the induced voltage in a sample with different $t_w$'s.
We obtained $F_s(t_w)$ by comparing $j_c$'s (with a criteria of the threshold voltage of 0.5 $\mu$V) for different $t_w$'s.
$I$-$V$ characteristics using short rectangular pulsed current were measured to estimate the total power, which causes the joule-heating effect.
$F_s(t_w)$ measurements were carried out below the critical value for heating.
Therefore, we safely concluded that the obtained $F_s(t_w)$ was not caused by the joule-heating effect in samples.
\begin{figure}[t]
\begin{center}
\includegraphics[width=0.99\linewidth]{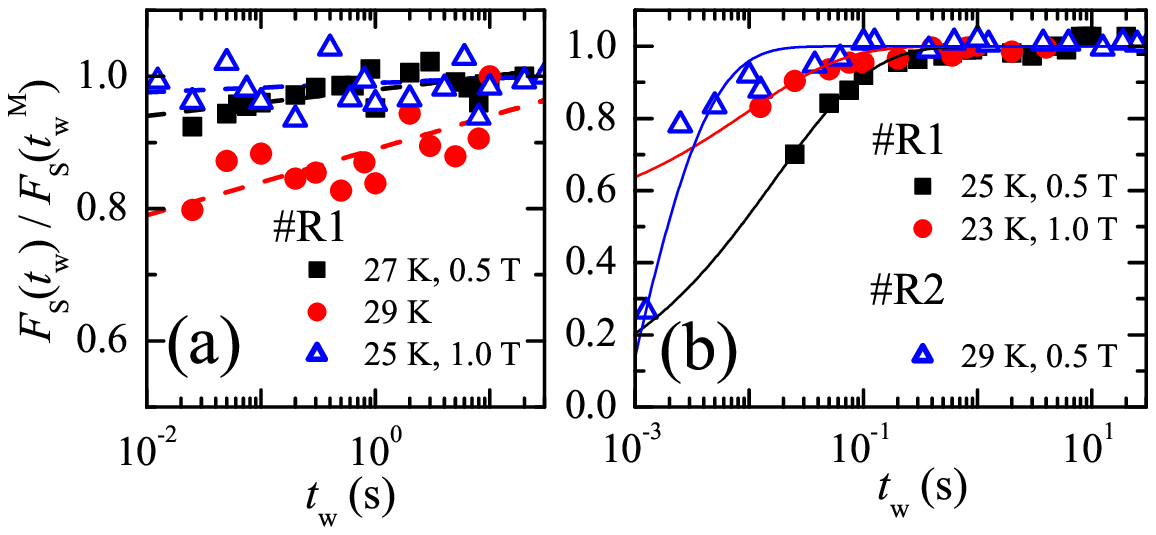}\\
\vspace{-1.2cm}
\includegraphics[width=0.99\linewidth]{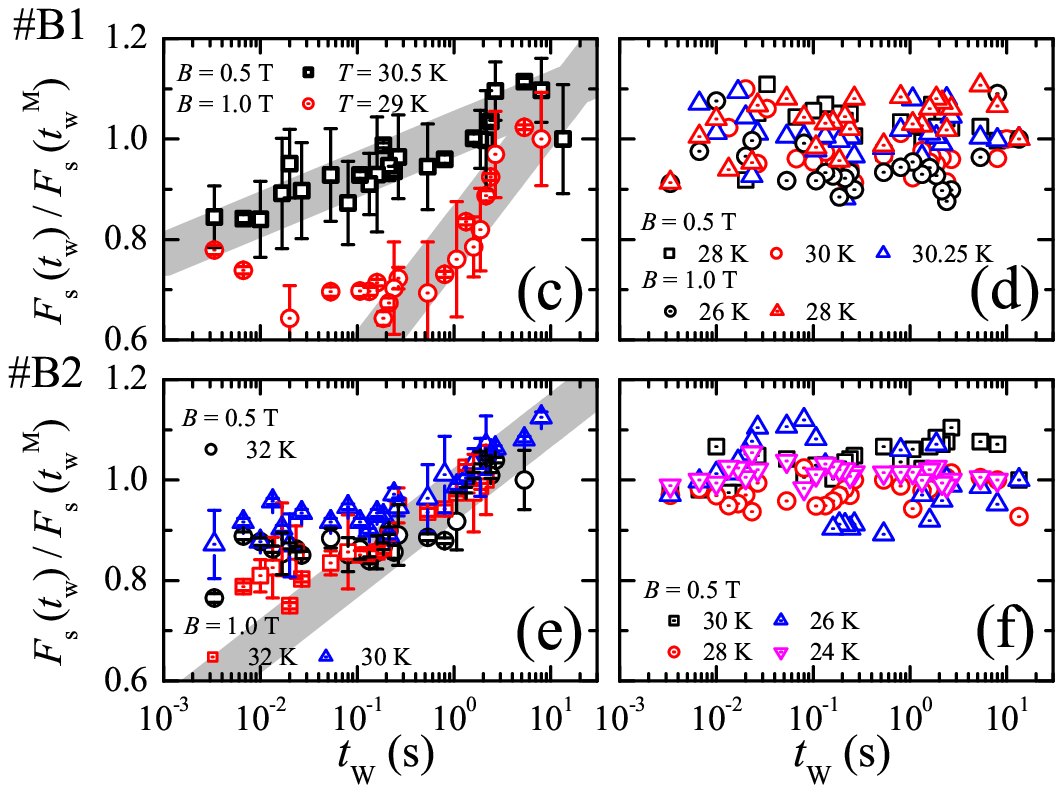}
\vspace{-1cm}
\caption{\label{label}The waiting-time dependence of the normalized maximum static friction force, $F_s(t_w) /  F_s (t_w^M)$, of (a)(b) samples with rectangular-type structure and (c)$\sim$(f) samples with bridge-type structure. The lines in (a) and (b) are the fitting curve using $F_s(t_w) \propto \log t_w$ and Eq. (2), respectively. The thick lines in (c) and (e) are guides to the eye.}
\end{center}
\end{figure}

We first discuss the results on samples with the rectangular-type structure (\#R1, \#R2).
Since $j_c$ strongly depends on temperature, we normalized the $j_c(t_w)$ data by $j_c$ at the maximum $t_w$, $t_w^M$.
Figure 2 shows the waiting-time dependence of the normalized maximum static friction force, $F_s(t_w) /  F_s (t_w^M)$.
We measured two different types of behaviour in the $B$-$T$ phase diagram.
When the temperature was close to the glass transition temperature, $T_g$\cite{Fisher-Fisher-Huse}, $F_s(t_w)$ depended logarithmically on $t_w$ as is shown in Fig. 2(a).
Typical values for the slope of the curve are 7$\times$10$^{-3}$ $\sim$ 5$\times$10$^{-2}$ s$^{-1}$.
On the other hand, with further decreasing temperature, $F_s(t_w)$ showed a rapid change for short $t_w$'s (Fig. 2(b)).
In other words, a characteristic time-scale, $t^\ast_w$ (typically 0.1 s at $B = $ 0.5 T and $T = $ 25 K in \#R1), showed up for the change of critical current at low temperatures.

For the results obtained above, we first argue the results in the high temperature region.
In the dynamics of driven vortices in high-$T_c$ superconductor, thermal fluctuation plays an essential role, and the so-called thermally assisted flux flow (TAFF)\cite{Anderson-Kim} leads to appreciable relaxation phenomena even at $j \ll j_c$.
Therefore, even after the driving force is stopped, the spatio-temporal profile of vortices in TAFF region can be described by a linear diffusion equation\cite{blatter}.
By solving this, it turned out that there is a characteristic time-scale $\tau \sim a_0^2 / D$, where $a_0$ is the intervortex distance ($\sim$ 50 nm) and $D$ is the diffusion coefficient of vortices.
If we take $D \sim 10^{-10}$ m$^2 /$s\cite{Moonen}, we obtain $\tau \sim 10^{-5}$ s.
Therefore, we expect that fast relaxation almost ceases for $t_w$ much longer than the above estimated $\tau$.
As a result, only very slow relaxation takes place for longer $t_w$.
Even though, we do not understand the reason for the logarithmic dependence.
However, this logarithmic waiting-time dependence can be also seen in the friction at the solid-solid interface\cite{scholz}.
Therefore, we can conclude that the essential physics of the logarithmic relaxation at the solid-solid interface is the same as that of the flux creep.

Next, we argue the results in the lower temperature region, where the strong $t_w$ dependence appeared below $t_w^\ast$.
First of all, it should be noted that the observed $t_w^\ast$ is much larger (4 orders of magnitude) than $\tau$ estimated in the high-temperature TAFF region.
It is almost impossible to explain such a huge difference after only 2 K temperature change (27 K and 25K at $B$ = 0.5 T in \#R1) in terms of the temperature dependence of the diffusion coefficient.
We can also consider the energy loss by the viscous motion of a single vortex during $t_w^\ast$ over distance $l$.
This should be equal to the depinning energy, such as $\eta \frac{l}{t^\ast _w} l = j_c \Phi_0 \xi$, where $\xi$ is the GL coherence length.
When we put $t^\ast _w = 0.1 $ s, $\eta = 5 \times 10^{-8} $ Ns/m$^2$\cite{Shibauchi}, $j_c = 1.75 \times $10$^3$ A/cm$^2$ and $\xi = 20 $ \AA, we obtain $l = 12 $ $\mu $m.
This is much larger than $a_0$($\sim $50 nm).
It is unlikely that a single vortex can move over such a huge distance ($l \gg  a_0$) without being pinned by the pinning center.

It is generally established that interaction between vortices becomes stronger at low temperatures, and  the vortex lattice stiffness increases and many vortices move coherently as a bundle\cite{bundle,Anderson-Kim}.
In case of the relaxation of vortex bundle, the above formula should be modified as $N\eta \frac{l}{t^\ast _w} l = j_c \Phi_0 \xi$, where $N$ is the number of vortex  in a bundle.
For $l \sim a_0$, $N \sim 6\times 10^4$, and the radius of the vortex bundle is about 12 $\mu$m.
Thus, it is very likely that large vortex bundles relax in a coherent manner.
The above estimate demonstrates that the correlation length of moving vortices becomes much larger than the static correlation length.
Such a long correlation length for moving vortex was also observed in the noise measurement of the bulk single crystal of Bi$_2$Sr$_2$CaCu$_2$O$_y$\cite{Maeda02}.
To be quantitative, using the collective creep theory, the data in Fig. 2(b) can be fitted by the following function\cite{Schnack},
\begin{equation}
 F_s(t_w) = \frac{F_s}{2} \left( 1+ \frac{1}{ [1+\frac{k_B T}{U_0}\ln (1+t_w / t_w^\ast)]^\frac{1}{\mu} } \right).
\end{equation}
Therefore, the relaxation observed at low temperatures suggests that the relaxation of a large vortex bundle did take place at lower temperatures.

It is interesting that Eq. (1) reminds us $F_s(t_w)$ of the boundary lubricated friction, where there exists very thin (2 or 3 atomic layers) glassy contamination material at the interface.
There the aging effect is described well by the following formula,
\begin{equation}
 F_s(t_w) = A + \frac{B\theta_0}{\theta_0 + (1-\theta_0) \exp [-(t_w/t_w^\ast)^{\beta}]},
\end{equation}
where $A, B, \theta_0$ and $ \beta$ are parameters, and $t_w^\ast$ gives the timescale with which the contamination material "solidifies" by the thermal re-distribution of molecules.
The data in Fig. 2(b) could be fitted well also by Eq. (2).
It should be noted that the $t_w$ dependence of Eq. (2) contains the stretched exponential form as the essential part, which is commonly observed in the relaxation of glassy systems.
Thus, these again suggest the importance of the relaxation of glassy movable object at low temperatures.


For the thin films with bridge-type structure (\#B1, \#B2), the logarithmic $F_s(t_w)$ is also observed close to $T_g$ (Figs. 2(c) and 2(e)), which indicates that the creep motion dominates the dynamics of vortices.
On the other hand, any relaxation can not be observed in both samples at lower temperatures (Figs. 2(d) and 2(f)).
The same feature was also found in \#B3 (not shown in the figure).
This effect was not observed in the rectangular-type samples, suggesting that this is due to the enhanced surface pinning at the edge of a sample.
The size of the vortex bundle increases with decreasing temperature.
When the size of the coherently moving vortex bundles becomes comparable to the size of the bridge region in a sample, the vortex bundles are effectively pinned by the edge of a sample\cite{Paltiel}, which hinders relaxations to take place.
We have not considered the spatial distribution of magnetic flux density explicitly, which has been discussed in terms of the critical state model, typically.
However, even if it is taken into account, essential features of the explanation presented above do not change.
These will be discussed in the separate publication\cite{myfuture}.

Moreover, we found that the boundary between the no-relaxation region and the logarithmic-relaxation region depends on the bridge width.
Figure 3(a) shows the "phase diagram" of the relaxation phenomena, in the reduced temperature ($t = T/T_g$) $vs$ bridge width ($w$) plane.
Boundary between the no-relaxation region (closed symbols) and the logarithmic-relaxation region (open symbols) is not vertical.
The thick line ($\propto t^{1.5}$) is a guide to the eye, which indicates that the smaller the sample size becomes, up to the higher temperatures the relaxation does not take place.
We can say that the behaviour in the smaller sample looks like the AC type friction in the sense that no relaxation takes place.
Therefore, it can be considered that one of the key component of AC law is the system size.


\begin{figure}[t]
\begin{center}
\begin{minipage}{0.45\linewidth}
\includegraphics[width=0.99\linewidth]{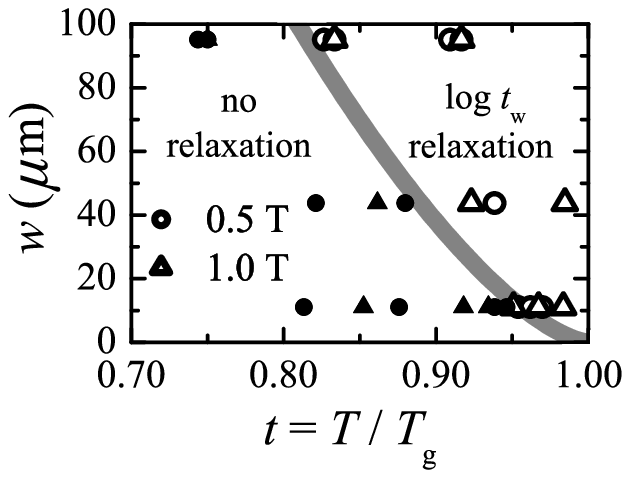}
\end{minipage}
\begin{minipage}{0.53\linewidth}
\includegraphics[width=0.8\linewidth]{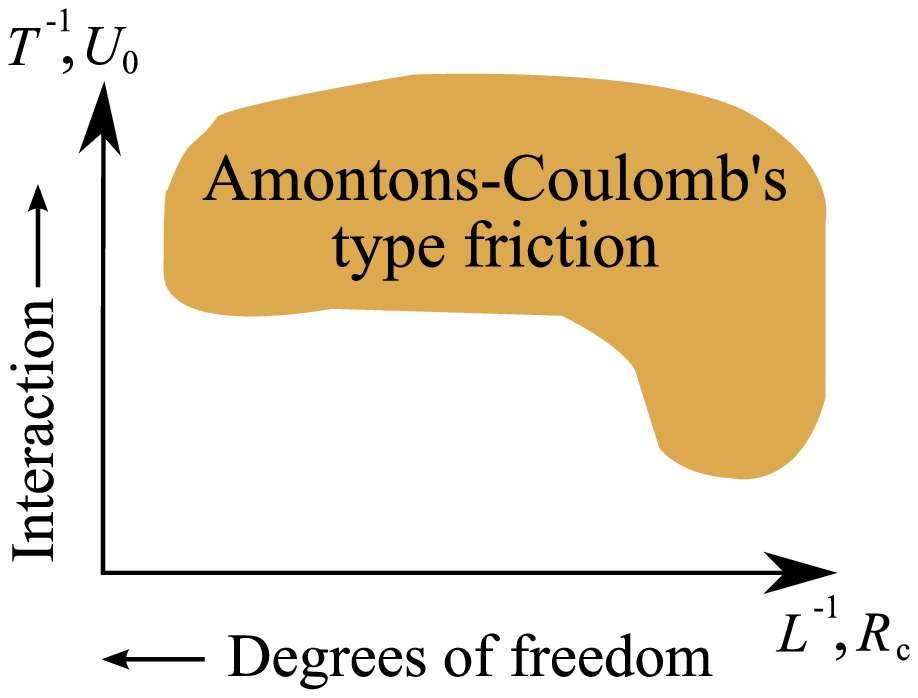}
\end{minipage}
\vspace{-0.2cm}
\caption{\label{label}(a)The size dependence of the relaxation in samples with bridge structure (\#B1$\sim$\#B3).
In the $w$-$t$ diagram, open symbols show the region $F_s(t_w) \propto \log t_w$, whereas closed symbols show the region $F_s(t_w) \sim $ const.
(b)Schematic figure on the validity of the AC type friction.} 
\end{center}
\end{figure}

Finally, we discuss the result on the bridge-type sample with columnar defects (\#C1).
For the heavy-ion irradiated sample, the drastic change for the relaxation was observed.
In this case, any relaxation cannot be observed in the $B$-$T$ phase diagram (not shown in the figure).
This is because vortices are trapped at the columnar defects.
In terms of the physics of friction, it is most like the AC type friction.
Therefore, together with the size effect, we can conclude that the strong pinning center leads to the AC type friction.


Based on these results, we try to deduce the criteria for the validity of the AC law.
The above results suggest that all of the pinning force, the thermal fluctuation, the size of coherently moving objects and the system size play crucial roles.
In short, the strong interaction force at the interface and the small degrees of freedom of the moving object are key aspects for the validity of the AC law.
These are drawn in a schematic figure, Fig. 3(b).
This is consistent with our previous results on the kinetic friction\cite{maedainoue}, where $F_k (v)$ is less velocity dependent with stronger pinning centers.
Therefore, there must be a universal parameter which discriminates the AC type friction from the all other types of the real friction.
Based on the results presented above, we believe that the parameter such as $R_c^{-1} \exp [ -U/k_B T_{\text{eff}}]$ can be a candidate for such a universal parameter, where $R_c$ is the radius of the coherently moving vortex bundle, $U$ is the activation energy, and $T^{\text{eff}} \propto T / L^3$ is the effective temperature, and $L$ is the system size.
More systematic study which changes the system size and the strength of the pinning force in more detail will clarify more quantitative aspects of our criteria.

In conclusion, the transient response of driven vortices was measured in La$_{2-x}$Sr$_x$CuO$_4$ thin films with different structures and pinning force, in terms of the physics of friction.
Using obtained results, we proposed a universal parameter which discriminates the AC type friction from all other types of the real friction.

We acknowledge C. J. van der Beek and M. Konczykowski for irradiation of columnar defects and fruitful discussions, S. Komiyama for supports in the instrumentation.
D. Nakamura also thanks to the Japan Society for the Promotion of Science for the financial support.


\end{document}